\documentclass{article}
\usepackage[T1]{fontenc}
\usepackage[utf8]{inputenc}
\usepackage[]{ismir} 
\usepackage{amsmath,url}
\usepackage{cite}

\usepackage{graphicx}
\usepackage{color}
\usepackage{xcolor}
\usepackage{amssymb}
\usepackage{pifont}
\usepackage{subcaption} 
\usepackage{array}
\usepackage{booktabs}
\usepackage{makecell}

\title{VioTech: Frame-Level Violin Playing Technique Detection with Multi-Scale Mixture-of-Experts}

\multauthor
  {Ting-Kang Wang$^{1,4}$ \hspace{1em} Vincent K.M. Cheung$^2$ \hspace{1em} Nan-Tien Lai$^3$ \hspace{1em} Li Su$^1$}
  {
  $^1$ Institute of Information Science, Academia Sinica, Taipei, Taiwan\\
  $^2$ Sony Computer Science Laboratories, Inc., Tokyo, Japan\\
  $^3$ National Taiwan University, Taipei, Taiwan\\
  $^4$ Original Content Center, Gamania Inc., Taipei, Taiwan \\
  {\tt\small tim.tkwang@gmail.com, cheung@csl.sony.co.jp, lisu@iis.sinica.edu.tw}
  }

\def\authorname{T.-K. Wang, V. K. M. Cheung, N.-T. Lai, and L. Su}

\usepackage[bookmarks=false,pdfauthor={\authorname},pdfsubject={\pdfsubject},hidelinks]{hyperref}

\begin{document}

\maketitle

\begin{abstract}

\noindent Playing technique is crucial for expressive violin performance, yet few violin technique detection models exist. Current systems mainly rely on note-level pipelines that first estimate note boundaries and then classify techniques. This is especially problematic for bowed string instruments such as the violin, where soft onsets and continuous bow-string interaction render note segmentation unreliable. Here, we propose \textbf{VioTech}, a frame-level violin technique detection framework that predicts techniques directly at 10 ms resolution without explicit note segmentation. To handle the large temporal-scale variation among techniques, VioTech combines a frozen MERT encoder with a multi-scale Mixture-of-Experts (MoE) architecture and task-specific routing for tonal, articulation, and legato-related predictions. We also introduce \textbf{VioTech120}, the first manually annotated dataset of violin playing techniques constructed from real solo performance recordings, as well as an accompanying note-labeling user interface. Results demonstrate that VioTech achieved consistent improvements over prior methods and significantly enhanced cross-dataset generalization, establishing a new state-of-the-art in violin technique-aware transcription.

\end{abstract}
\begin{figure}
  \centering
  \includegraphics[width=1\linewidth]{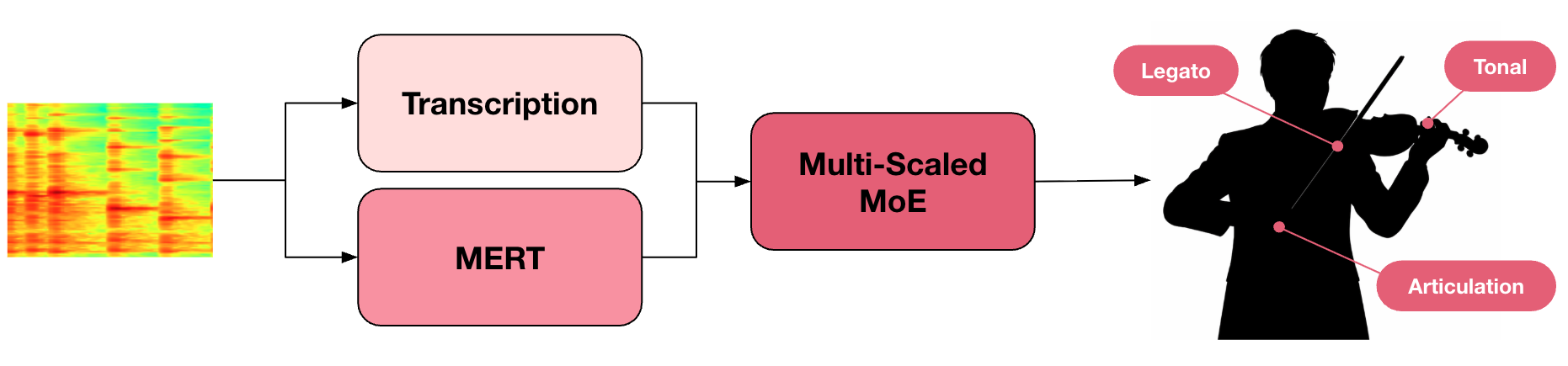}
  \caption{Overview of proposed violin technique transcription system with multi-scale mixture of experts.}
  \label{fig:overview}
\end{figure}

\section{Introduction}\label{sec:introduction}

Violin playing techniques such as \textit{spiccato}, \textit{pizzicato}, \textit{legato}, and \textit{flageolet}, are central to expressive performance, yet their automatic detection from audio remains challenging, particularly at the frame level. Existing approaches typically adopt a note-level pipeline, where note boundaries are first estimated and techniques are then classified per note~\cite{11464063}. This design is however fragile, as classification performance depends on accurate note segmentation that is unreliable for bowed string instruments due to soft onsets and continuous bow–string interaction.

Frame-level prediction offers a principled alternative by removing the need for explicit segmentation. However, applying frame-level modeling to violin technique detection introduces a new challenge: techniques exhibit inherently multi-scale temporal characteristics. For example, pizzicato is localized to a short attack transient, spiccato depends on the bowing envelope, and bow-change events require longer, cross-note contexts. As a result, no single temporal window can adequately capture all relevant temporal dependencies.

To overcome these challenges, we propose \textbf{VioTech}, a frame-level violin playing technique-aware transcription system designed to address the inherently multi-scale nature of violin playing techniques. To this end, VioTech adopts a multi-scale mixture-of-experts (MoE) architecture that integrates complementary acoustic and note-aware representations. A transcription backbone provides frame-level note-aware cues, while a frozen acoustic encoder supports a set of temporal experts operating at different time scales. Through adaptive routing, the model captures technique-specific temporal dependencies while remaining interpretable across temporal and acoustic dimensions.

To support this task, we further introduce \textbf{VioTech120}, the first dataset of violin playing techniques constructed from real solo performances. The dataset comprises 120 manually annotated violin pieces, including technically demanding repertoire such as \emph{Paganini 24 Caprices}.  VioTech120 provides fine-grained note-level technique annotations rather than coarse file-level labels, covering six techniques and including, to our knowledge, the first explicit annotation of legato events. This design enables precise and diverse supervision for fine-grained technique recognition, with the accompanying note-labeling user interface  enabling additional data generation with precise temporal control.

Our contributions are summarized as follows:
\begin{itemize}
    \item We propose \textbf{VioTech}, a frame-level violin technique detection framework that eliminates explicit note segmentation and addresses the multi-scale nature of violin techniques through a Mixture-of-Experts architecture. We show that VioTech achieved excellent transcription performance and generalization.

    \item We introduce \textbf{VioTech120}, the first manually annotated dataset of violin playing techniques constructed from real solo performance recordings, providing fine-grained annotations.

    \item We release a piano-roll-based annotation interface that enables efficient and consistent note-level labeling of violin techniques, with precise audio-aligned temporal control that facilitates fine-grained frame-level modeling and dataset construction.
\end{itemize}

The dataset, codebase, and annotation interface are available on our project page.\footnote{Project page: \url{tklovln.github.io/VioTech/}}


\begin{figure*}[t]
  \centering
  \includegraphics[width=.9\textwidth]{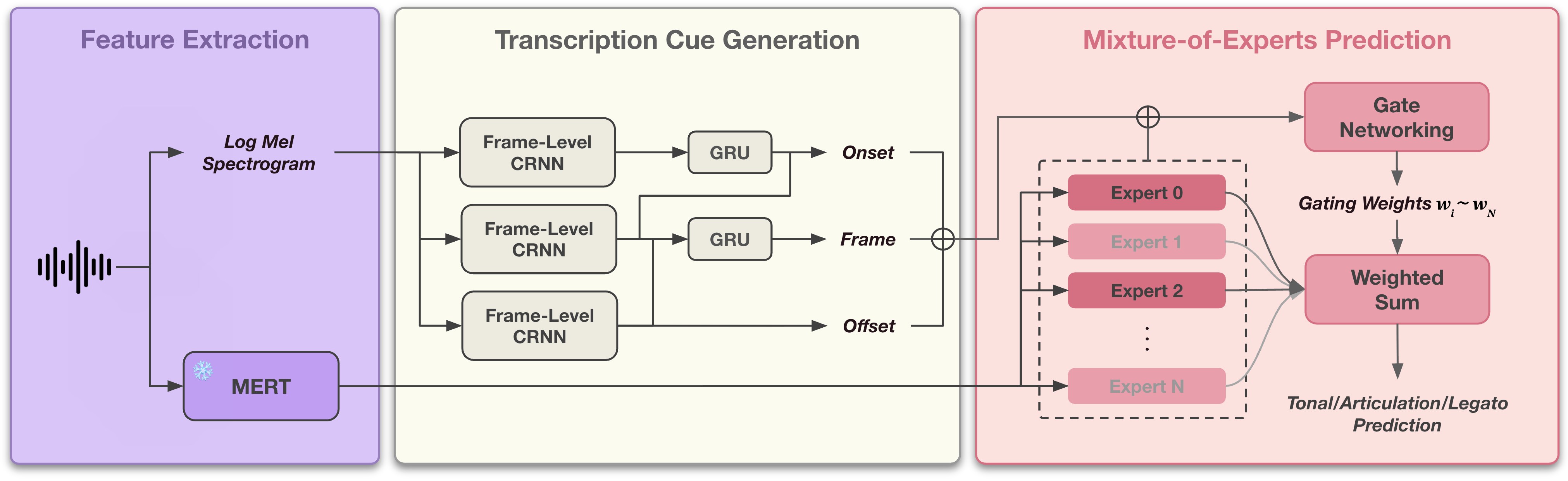}
  \caption{Proposed multi-scale mixture-of-experts model architecture for frame-level violin technique detection.}
  \label{fig:architecture}
\end{figure*}

\section{Related Work}\label{sec:related work}

\subsection{Violin Playing Technique Classification}
Early work on violin technique recognition mainly framed the task as isolated-note or pre-segmented clip classification. Using temporal, spectral, cepstral, harmonic, and phase-derived features, Su et al. \cite{su2014sparse} showed that sparse coding of magnitude and phase-derived spectra were most effective technique classification. Kruger et al. \cite{kruger2020playing} later combined traditional audio features with Hartley-transform-based features using SVM classifiers. Alar et al.\ \cite{alar2021audio} further explored CNN-based classification from MFCC features. While these studies demonstrate the feasibility of violin technique classification, they largely assume isolated notes or fixed-duration inputs. Recently, Wang et al.\ \cite{11464063} proposed VioPTT, a note-level violin technique-aware transcription system using synthetic MIDI-augmentations. However, that design depended heavily on accurate note boundary detection, which remains a known bottleneck for transcription pipelines~\cite{duxbury2002hybrid, tomczak2023onset}.

\subsection{Self-Supervised Audio Representations for Music}
Self-supervised learning has significantly advanced music representation learning. Examples such as MERT \cite{li2023mert} that allow for low-level acoustic to higher-level semantic abstractions have been used to fine-tune entire model backbones or extract single-pooled representations for downstream tasks such as audio tagging and music classification \cite{li2023mert, yuan2023marble}. Recently, MERTech\cite{li2024mertech} applied MERT to instrument playing technique detection through task-specific finetuning. It combines technqiue prediction with auxiliary tasks such as pitch and onset detection to improve robustness under limited data and class imbalance. This suggests that MERT-based representations can be effectively adapted to fine-grained technique modeling tasks.

\subsection{Multi-Scale Temporal Modeling}
Relatedly, multi-scale temporal modeling has been extensively applied to audio and music processing. Dilated convolution stacks as used in WaveNet-style architectures \cite{oord2016wavenet} expand receptive fields without a proportional increase in parameters. Feature pyramid networks aggregate representations at multiple resolutions for detection tasks \cite{lin2017feature}. Convolution-based systems and attention-based pooling mechanisms have been further used to capture events of varying duration \cite{cakir2017convolutional, kong2020sound} in frame-level sound event detection as well as timbral and temporal structure in MIR tasks \cite{pons2017timbre, won2020evaluation}. 

\subsection{Mixture-of-Experts Mechanism}
While multi-scale modeling allows for abstraction across different timescales, information is usually combined within shared representations. To make temporal-scale specialization explicit and task-dependent, as well as to enhance interpretability, we therefore turn to a mixture-of-experts (MoE). The use of MoE for fine-grained music analysis has so far been limited. However, MoE architectures have been widely used in natural language processing where a gating network selects experts based on input features. This enables the handling of heterogeneous acoustic conditions such as noise environments or speaker characteristics in speech recognition \cite{shazeer2017outrageously}, as well as conditional computation and specialization in Switch Transformer \cite{fedus2022switch} and GShard \cite{lepikhin2020gshard}.

\section{Methodology}\label{sec: methodology}

Our system consists of two components: (1) a transcription module that produces note-aware cues, and (2) a multi-scale MoE module that integrates these cues with representations from a frozen MERT model for technique prediction. An overview is shown in Figure~\ref{fig:architecture}.

\subsection{Transcription Cue Generation}

We adopt the High-Resolution Piano Transcription architecture \cite{kong2021high}, repurposed for violin in \cite{11464063}. The input is a mono waveform at 16 kHz. A short-time Fourier transform followed by a log-mel filterbank produces a spectrogram at 100 frames per second. 

After batch normalization, three parallel branches predict frame activity, onset, and offset, each producing a tensor of shape $(B, T, 88)$ corresponding to batch size, number of frames, and MIDI pitches 21--108. Each branch consists of convolutional blocks followed by a fully connected layer, a bidirectional GRU, and a linear projection with sigmoid activations for multi-label prediction. A GRU-based fusion stage refines the outputs by conditioning frame predictions on onset and offset signals. The outputs of these three branches are used as transcription cues to provide note-aware context for the subsequent MoE architecture. 


\subsection{Multi-Scale Mixture-of-Experts}

\begin{figure}[t]
\centering
\newlength{\panelheight}

\setlength{\panelheight}{5cm}

\begin{subfigure}[t]{0.43\linewidth}
    \centering
    \begin{minipage}[c][\panelheight][c]{\linewidth}
        \centering
        \small
        \renewcommand{\arraystretch}{1.4}
         \resizebox{\linewidth}{!}{%
        \begin{tabular}{|c|c|c|}
        \hline
        Expert & $(K, D)$ & \makecell{Window size} \\
        \hline
        0 & $(2, 1)$ & 20ms\\
        1 & $(4, 1)$ & 40ms\\
        2 & $(8, 1)$ & 80ms\\
        3 & $(16, 1)$ & 160ms\\
        4 & $(32, 1)$ & 320ms\\
        5 & $(64, 1)$ & 640ms\\
        6 & $(128, 1)$ & 1280ms\\
        7 & $(256, 1)$ & 2560ms\\
        \hline
        \end{tabular}
        }
    \end{minipage}
    \caption{}
    \label{fig:expert_rf}
\end{subfigure}
\hfill
\begin{subfigure}[t]{0.53\linewidth}
    \centering
    \begin{minipage}[c][\panelheight][c]{\linewidth}
        \centering
        \includegraphics[height=\panelheight,width=\linewidth,keepaspectratio]{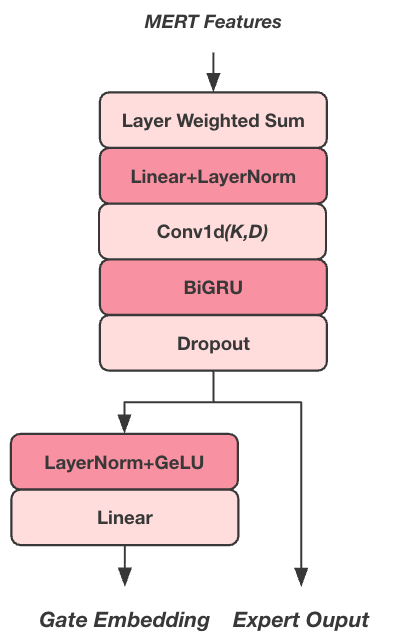}
    \end{minipage}
    \caption{}
    \label{fig:expert_arch}
\end{subfigure}

\caption{(a) Configuration and corresponding temporal context window size of our multi-scale mixture-of-experts with convolution kernel size $K$ and dilation size $D$. (b) Architecture of a single expert.}
\label{fig:expert_combined}
\end{figure}

\subsubsection{Multi-Scale Temporal Experts}

We build $N$ temporal experts on top of a frozen MERT backbone. Given an input waveform, MERT-v1-330M produces $L+1$ hidden-state tensors (here $L= 25$), each with shape $(B, T_{\text{mert}}, 1024)$. Each expert $i$ first learns a set of layer weights $\mathbf{w}_i \in \mathbb{R}^{25}$ from each hidden layer $\mathbf{H}_l$ to obtain the representation
\[
\mathbf{h'}_i = \sum_{l=0}^{24} \text{softmax}(\mathbf{w}_i)_l \cdot \mathbf{H}_l,
\]
for which the resulting features are projected to 128 dimensions and temporally aligned with the transcription frames.

Temporal context is next modeled using a depthwise separable convolution with kernel size $k_i$ and dilation $d_i$, followed by pointwise convolution, normalization, and GELU activation. This defines a temporal context window of $k_i \cdot d_i \cdot 10$ ms, ranging from 20 ms to 2560 ms (see Figure \ref{fig:expert_combined}a). Kernel sizes are exponentially spaced, allowing experts to specialize at different temporal scales.

Finally, features are processed by an independent bidirectional GRU to produce content representations $\mathbf{h}_i \in \mathbb{R}^{(B, T, 256)}$, along with a separate projection to gate summaries $\mathbf{g}_i$. The architecture of a single expert is shown in Figure~\ref{fig:expert_arch}.

\subsubsection{Per-Task Gating and Classification Heads}

This stage aggregates expert outputs and produces task-specific predictions. The onset, offset, and frame outputs from the transcription backbone are projected to 16-dimensional embeddings and concatenated to form a cue vector $\mathbf{c}$. For each expert $i$, the content feature $\mathbf{h}_i$ is concatenated with $\mathbf{c}$ and passed through a two-layer MLP to produce $\mathbf{e}_i \in \mathbb{R}^{(B, T, 128)}$. Stacking across experts yields $\mathbf{E} \in \mathbb{R}^{(B, T, N, 128)}$. Gate summaries from all experts are then concatenated with $\mathbf{c}$ to form $\mathbf{g}_{\text{in}}$. Three task-specific gating networks (tonal, articulation, legato) map $\mathbf{g}_{\text{in}}$ to softmax weights $\boldsymbol{\alpha}^{\text{task}} \in \mathbb{R}^N$, which are used to combine expert features:
\[
\mathbf{m}^{\text{task}} = \text{ReLU}\bigl(\text{LayerNorm}\bigl(\sum_i \alpha_i^{\text{task}} \cdot \mathbf{e}_i\bigr)\bigr).
\]

The final predictions are produced by three independent heads. The tonal head outputs three classes (normal, pizzicato, flageolet), the articulation head outputs two classes (normal, spiccato), and the legato head outputs a scalar probability via sigmoid. Legato is formulated as a regression task with Gaussian targets centered at bow change points, enabling continuous estimation of bow transitions.



\section{VioTech120 Dataset}

\subsection{Technique Annotation}
\begin{figure}[t]
    \centering
    \includegraphics[trim={0.5cm 0cm 0.5cm 2.5cm}, clip, width=1.\linewidth]{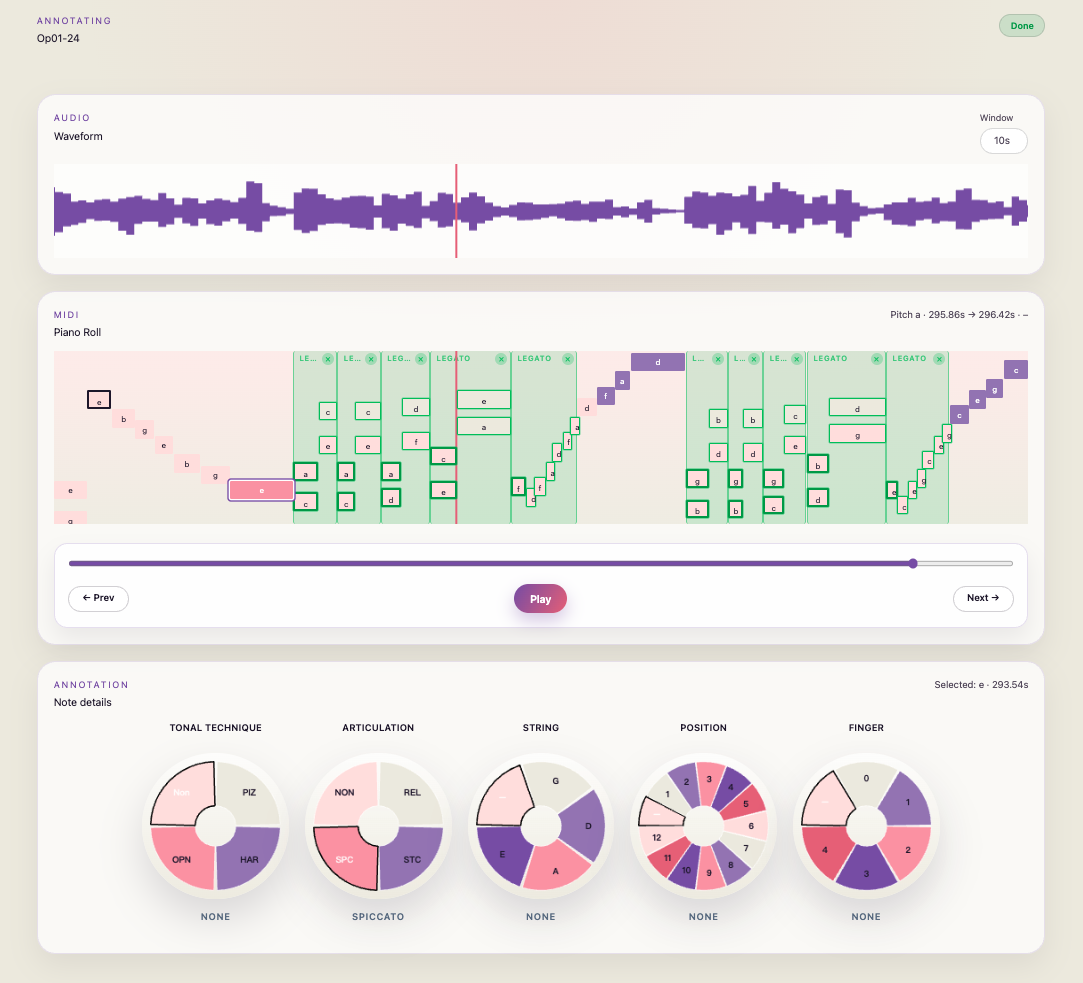}
    \caption{We provide a user-friendly annotation interface with fine-grained temporal alignment and control between MIDI and audio to enable data extension beyond playing technique and include additional fingering information.}
    \label{fig:annotation_interface}
\end{figure}

The VioTech120 dataset consists of 120 solo violin etudes with aligned audio and MIDI from three standard collections: 24 caprices from \emph{Paganini, Op.\ 1}, 60 studies from \emph{Wohlfahrt, Op.\ 35}, and 36 studies from \emph{Kayser, Op.\ 20} based on the ViolinMIDIDataset~\cite{tamer2023high}. Technique labels are annotated at the note level using aligned MIDI as the temporal reference. Each note is assigned labels along three dimensions: tonal, articulation, and legato. For tonal techniques, we use Tonal Normal (TN), Tonal Flageolet (TF), and Tonal Pizzicato (TP). For articulation, we use Articulation Normal (AN) and Articulation Spiccato (AS). For legato, we model bow-change events using soft frame-level targets, where each annotated bow-change onset is represented by a Gaussian peak centered at the onset frame. This yields a probability-like target in which values close to 1 indicate bow-change frames (LB), while values close to 0 correspond to sustained frames (LS). For brevity, we use these abbreviations throughout the paper. These dimensions are not mutually exclusive: Labels are assigned in a multi-label manner, allowing multiple technique attributes to co-occur on the same note.

The annotation was carried out by two semi-professional violinists with over ten years of formal training, and all labels are finalized through a peer review process to ensure consistency. 


\subsection{Annotation Interface}
To support efficient and consistent annotation, we developed a user annotation interface based on a piano-roll visualization as shown in Figure~\ref{fig:annotation_interface}. The interface allows direct interaction with the audio signal while providing an aligned view of MIDI notes, enabling annotators to assign technique labels with precise temporal control effectively and with ease. In addition to technique labels, the interface supports annotation of violin position and fingering, which facilitates future extensions of the dataset and improves usability for annotators.

\section{Experimental Setting}\label{sec: experiments}

\subsection{Dataset}

In addition to the curated VioTech120 dataset, we incorporate MOSA-VPT~\cite{11464063}, a large-scale synthetic dataset of violin playing techniques. MOSA-VPT contains approximately 76 hours of audio-MIDI pairs, balanced across four techniques: detaché, flageolet, spiccato, and pizzicato. Although this dataset provides only file-level technique labels, it includes a substantial number of samples for relatively scarce techniques such as flageolet and pizzicato, which complements the manually annotated VioTech120 dataset. During training, VioTech120 and MOSA-VPT data were combined at an 8:2 ratio, and the validation set followed the same proportion.

For cross-dataset evaluation, we used the RWC (Real World Computing) Music Database~\cite{goto2003rwc} as test data. Only violin recordings were evaluated and scales were first segmented into individual notes following prior work \cite{su2014sparse, 11464063}.

\subsection{Training Details}

We initialized all models with a transcription backbone pretrained on VioPTT \cite{11464063} and kept the MERT-v1-330M encoder frozen. Models were trained using Adam with a learning rate of $5\times10^{-4}$ and batch size of 8. Learning rate decay and early stopping were applied based on validation performance, and the best checkpoint was selected with technique prediction macro F1. For training, we used 10-second audio segments and the same standard audio augmentations in VioPTT \cite{11464063} that included pitch shifting, gain boost, reverb shifting, and random band-pass filtering.

The transcription branch was trained with regression losses for onset and offset and binary cross-entropy for frame prediction. For technique prediction, tonal and articulation tasks used cross-entropy with focal loss, together with a load-balancing regularization term. The total loss is given by
\[
\mathcal{L}_{\text{total}} = \mathcal{L}_{\text{transcription}} + \mathcal{L}_{\text{tech}},
\]
where $\mathcal{L}_{\text{tech}}$ aggregates all technique losses.

\begin{figure*}[t]
\centering
\setlength{\panelheight}{7.2cm} 

\begin{minipage}[c]{0.56\textwidth}
\centering

\renewcommand{\arraystretch}{1.2}
\resizebox{\linewidth}{!}{%
\begin{tabular}{lcccccc}
\hline
\textbf{MODEL} 
& \textbf{TN $\uparrow$} 
& \textbf{TP $\uparrow$} 
& \textbf{TF $\uparrow$} 
& \textbf{AN $\uparrow$} 
& \textbf{AS $\uparrow$} 
& \textbf{MACRO $\uparrow$} \\
\hline
VioTech (Ours)
& 79.07 & 89.60 & \underline{88.84} & 97.78 & \textbf{97.16} & \textbf{90.49} \\
w/o MERT Layer Weight
& \textbf{86.11} & \underline{93.68} & 65.44 & \underline{98.74} & 83.66 & \underline{85.53} \\
w/o Transcription Cue
& 79.83 & \textbf{95.28} & 83.28 & \textbf{99.57} & 64.51 & 84.49 \\
w/o MoE 
& \underline{85.98} & 40.85 & 69.62 & 94.43 & 61.41 & 70.46 \\
\hline
VioPTT~\cite{11464063} 
& 77.22 & 88.80 & 71.89 & 77.22 & \underline{85.08} & 80.04 \\
MERTech~\cite{li2024mertech} 
& 53.36 & 43.27 & \textbf{95.77} & 53.36 & 15.61 & 52.27 \\
\hline
\end{tabular}%
}
\captionof{table}{Technique detection accuracy (\%) on the RWC test set, excluding legato for comparison with prior work. Best results are in bold, and second best are underlined.}
\label{tab:per_class_rwc_full}

\vspace{1.0em}

\renewcommand{\arraystretch}{1.2}
\resizebox{\linewidth}{!}{%
\begin{tabular}{lcccccccc}
\hline
\textbf{MODEL} 
& \textbf{TN $\uparrow$} 
& \textbf{TP $\uparrow$} 
& \textbf{TF $\uparrow$} 
& \textbf{AN $\uparrow$} 
& \textbf{AS $\uparrow$}  
& \textbf{LS $\uparrow$} 
& \textbf{LB $\uparrow$} 
& \textbf{MACRO $\uparrow$} \\
\hline
VioTech (Ours)
& \underline{87.40} & \underline{94.00} & 30.10 & \textbf{98.69} & \textbf{90.50} & 97.61 & 40.02 & \textbf{76.90} \\
w/o MERT Layer Weight
& \textbf{91.28} & 85.35 & \textbf{35.29} & \underline{98.34} & \underline{86.46} & 97.89 & 43.42 & \underline{76.86} \\
w/o Transcription Cue
& 88.09 & \textbf{95.59} & 30.03 & 97.58 & 76.86 & \underline{97.98} & \underline{47.17} & 76.19 \\
w/o MoE
& 88.38 & 57.46 & \underline{30.87} & 94.72 & 59.78 & \textbf{98.92} & \textbf{64.09} & 70.60 \\
\hline
\end{tabular}%
}
\captionof{table}{Technique detection F1 score (\%) on the RWC test set with legato included.}
\label{tab:per_class_f1_rwc_ablation}

\end{minipage}
\hfill
\begin{minipage}[c]{0.42\textwidth}
\centering

\includegraphics[
    width=\linewidth,
    height=0.47\panelheight,
    keepaspectratio,
    trim={0cm 0cm 0cm 0.9cm},
    clip
]{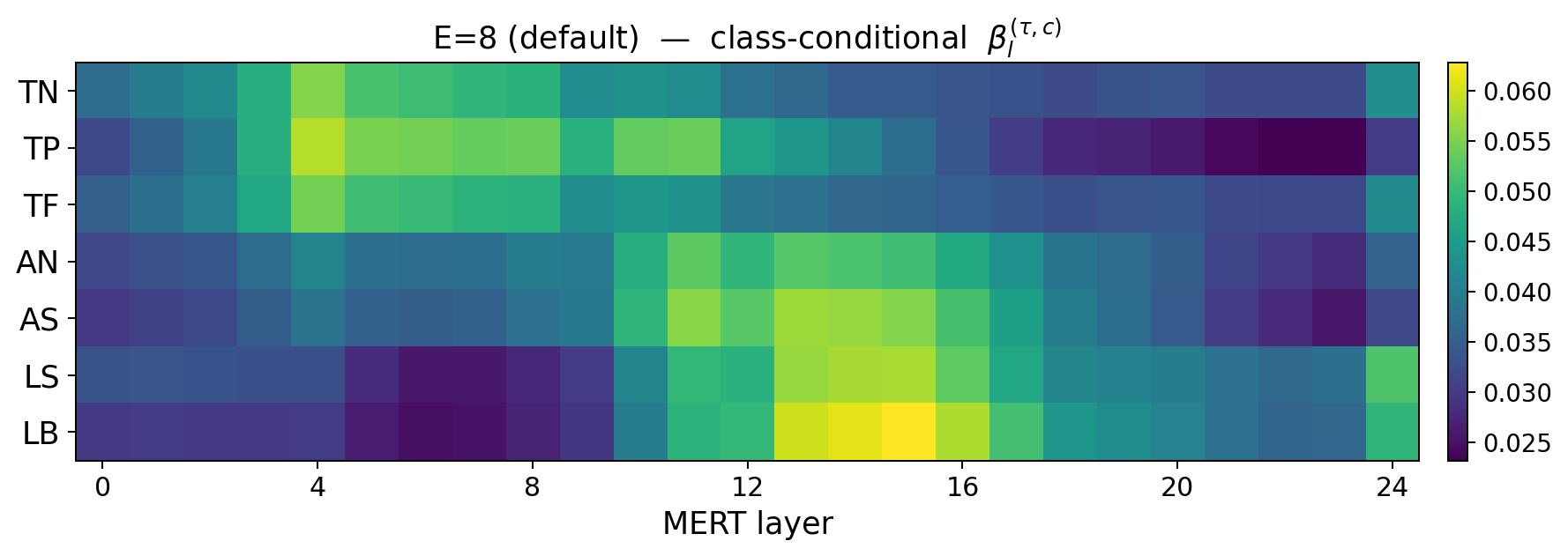}

\vspace{0.4em}

\includegraphics[
    width=\linewidth,
    height=0.60\panelheight,
    keepaspectratio,
    trim={0cm 0cm 0cm 0.9cm},
    clip
]{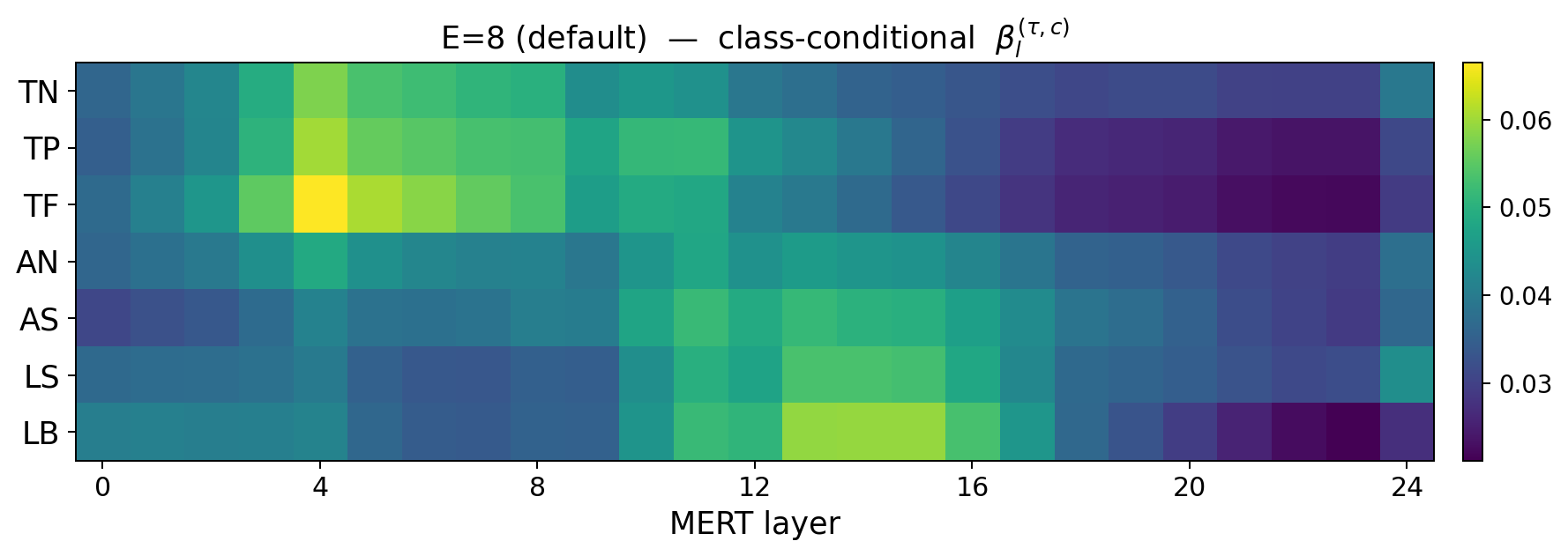}

\captionof{figure}{Learned MERT layer weights by violin playing technique when evaluated on VioTech120 test (top) and RWC (bottom) datasets.}
\label{fig:beta_class}

\end{minipage}

\end{figure*}

\section{Results And Discussion}

\subsection{Frame-Level Technique Detection Performance}

We first examined technique detection accuracy in Table~\ref{tab:per_class_rwc_full} to compare with prior work that does not include legato annotations. Table~\ref{tab:per_class_f1_rwc_ablation} further reports F1 scores with legato labels included. 

In the shared-label accuracy setting, VioTech achieved the highest macro accuracy of 90.49\%, outperforming VioPTT (80.04\%) and MERTech (52.27\%). The improvement was especially clear for articulation-related techniques, where VioTech obtained the highest spiccato accuracy (AS, 97.16\%), while maintaining competitive performance on tonal techniques such as pizzicato (TP) and flageolet (TF). Under the more-stringent F1-based full-task evaluation, VioTech also achieved the best overall macro F1 of 76.90\%, indicating that its advantage was not limited to accuracy-based evaluation. 

Compared with the non-MoE variant, we further observe that the proposed multi-scale MoE method substantially improved transient and articulation-related techniques: pizzicato F1 increased from 57.46\% to 94.00\%, and spiccato F1 increased from 59.78\% to 90.50\%. These gains suggest that multi-scale expert routing effectively captured short-term temporal dynamics and attack characteristics, which are critical for transient violin techniques. 

However, bow-change detection (LB) decreased from 64.09\% in the non-MoE baseline to 40.02\% in VioTech. We attribute this drop to the sparse and long-range nature of bow-change events, which require extended temporal context and may be under-emphasized when expert routing favors shorter temporal scales. This reveals a trade-off in the current multi-scale design: expert specialization improves transient-event detection but remains less effective for sparse phrase-level phenomena. 

Component ablations further show that removing the technique weighting had little effect on macro F1 (76.90\% vs.\ 76.86\%), suggesting that temporal expert specialization contributed more than layer-wise feature selection. Removing transcription cues led to a moderate decrease in macro F1 (76.19\%), indicating that note-aware information provided complementary benefits. Finally, flageolet (TF) remained difficult to transcribe across all configurations, with relatively low F1 despite high accuracy, likely due to its subtle acoustic characteristics and low prevalence at the frame level. 

These results show that VioTech improved cross-dataset generalization mainly by enhancing the detection of transient and articulation-related techniques, while also highlighting the need for long-context experts for bow-change modeling in future designs.

\subsection{Note Transcription}

\begin{table}[t]
\centering

\resizebox{\columnwidth}{!}{%
\begin{tabular}{lcccc|cccc}
\toprule

\textbf{Model} 
& \multicolumn{4}{c}{\textbf{Bach10}} 
& \multicolumn{4}{c}{\textbf{URMP}} \\

\midrule
&  P & R & F1 & F1$_\text{w/o}$
& P & R & F1 & F1$_\text{w/o}$ \\


\cmidrule(lr){2-5} \cmidrule(lr){6-9}

VioPTT \cite{11464063}
& 48.7 & 53.0 & 50.7 & 62.5
& 44.4 & 44.9 & 44.6 & 54.9 \\

VioTech w/o MoE
& 54.8 & 58.0 & 56.1 & 65.2
& 39.1 & 57.5 & 46.3 & 59.7 \\

VioTech (Ours)
& \textbf{60.9} & \textbf{65.0} & \textbf{62.8} & \textbf{79.1}
& \textbf{45.6} & \textbf{60.4} & \textbf{51.9} & \textbf{69.3} \\

\bottomrule
\end{tabular}%
}
\caption{Note transcription performance on Bach10 \cite{datasetbach10} and URMP \cite{li2018creating} datasets. P = precision, R = recall, F1$_\text{w/o}$ = F1-score without note offsets.}
\label{tab:transcription_main}
\end{table}

Although note transcription is not the primary objective of VioTech, it consistently outperformed prior technique-aware transcription models on Bach10~\cite{datasetbach10} and URMP~\cite{li2018creating} (Table~\ref{tab:transcription_main}), evaluated with \texttt{mir\_eval}~\cite{raffel2014mir_eval} following the setting in VioPTT~\cite{11464063}. The consistent gain over the non-MoE variant further indicates that multi-scale expert modeling benefits both technique prediction and note transcription.

\subsection{Scaling Behavior of Temporal Experts}
\begin{figure*}[t]
    \centering
    \includegraphics[width=1\linewidth,trim={0cm 0cm 0cm 0.8cm}, clip]{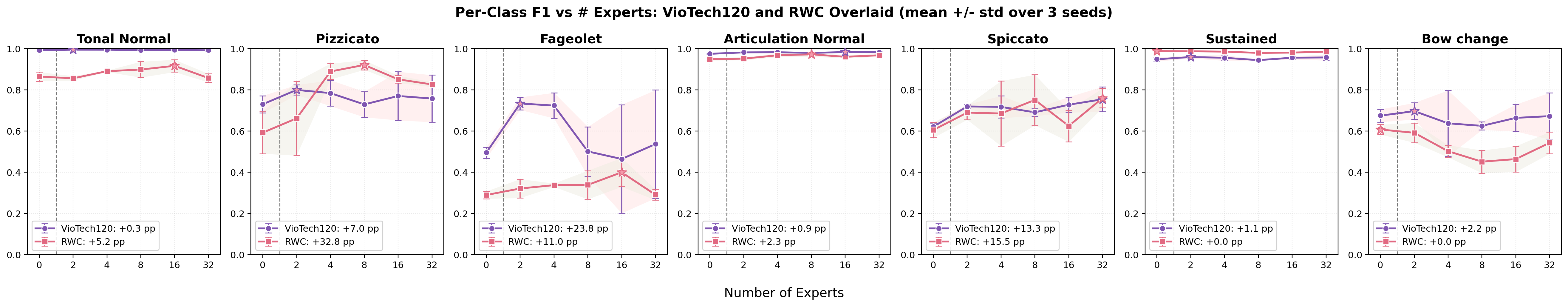}
    \caption{
    Technique-wise F1-score versus number of experts evaluated on VioTech120 test and RWC \cite{goto2003rwc} datasets. Stars indicate highest mean performance over three iterations and error bars indicate standard deviation.
    \emph{pp} denotes percentage-point improvement over the no-MoE baseline.
    }
    \label{fig:per_class_f1_experts}
\end{figure*}

\begin{figure}[b]
    \centering
    \includegraphics[
    width=0.95\linewidth,
    trim=0 0cm 0 2.3cm,
    clip
    ]
    {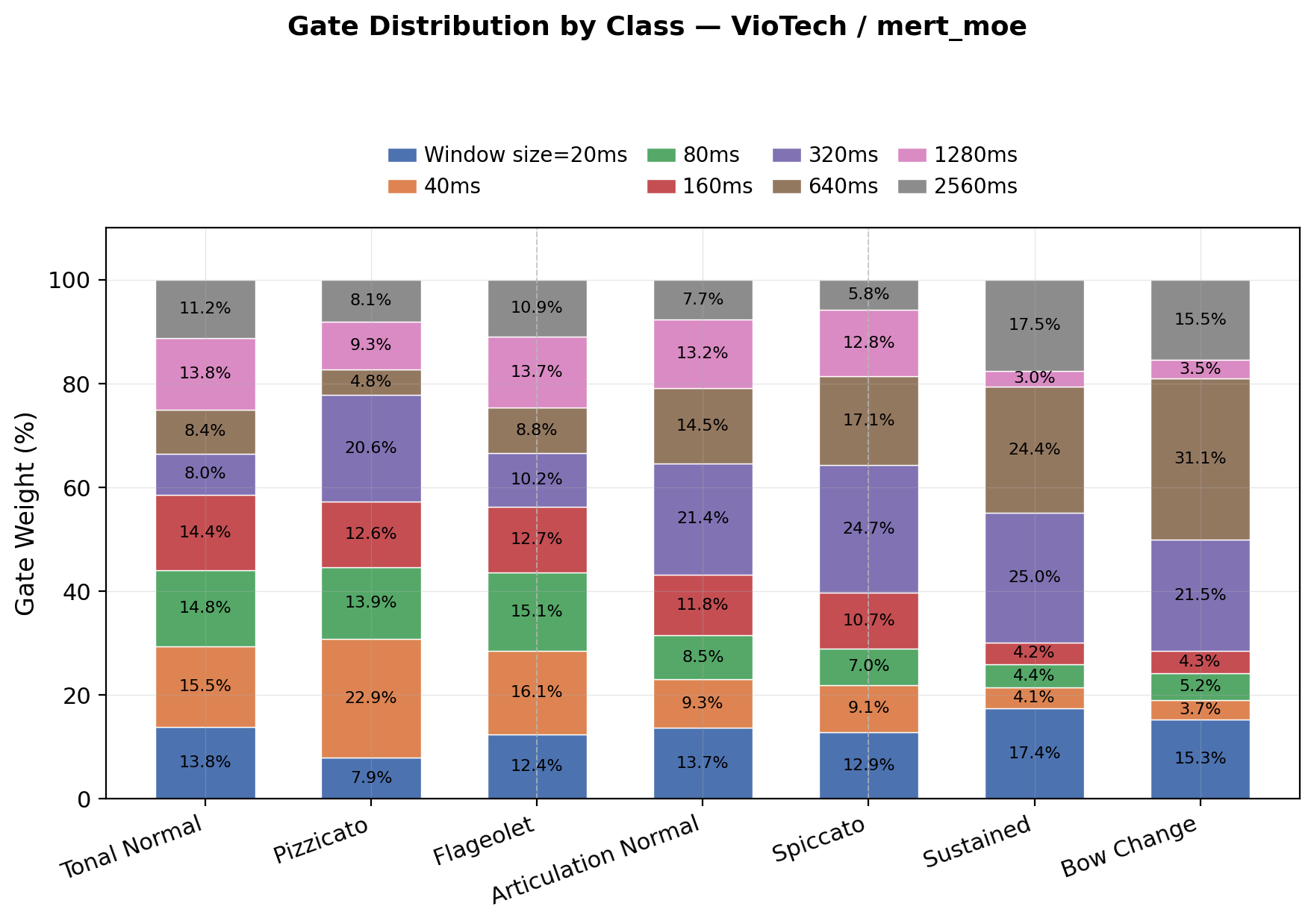}
    \caption{Gate distribution across experts for each violin playing technique.}
    \label{fig:gate_class}
\end{figure}

Here, we tested how the granularity of the temporal context window of each expert affected detection performance on the single-technique level. We varied the number of experts between $2^i$ with $i=1,2,3,4,5$, for which the kernel size/temporal context window was evenly divided on the exponential scale (see Figure \ref{fig:expert_combined}a). Results were compared against our model without any experts (hence shared representations across multiple scales) as baseline. This experiment was repeated three times for consistency.

We observe from Figure~\ref{fig:per_class_f1_experts} that the MoE approach provided significant performance gains on expressive and less frequent techniques, such as pizzicato and spiccato, especially when evaluated cross-dataset on RWC. This suggests that expert routing helps assign temporal modeling capacity to class-specific acoustic patterns that are difficult to capture with a single shared head. Nevertheless, MoE only provided marginal improvements for frequent classes such as normal or sustained, whose performance was already near ceiling. 

However, more experts did not always equate to better performance: the number of experts that enabled the highest F1 scores differed across techniques and datasets. For example, a jump in performance was seen for flageolet with two and four experts on VioTech with decreased mean performance and increased variability with increased number of experts, whereas the best performance was obtained with eight experts on RWC.

Taken together, while the proposed MoE approach improves technique detection performance and interpretability, the optimal number of experts and consequent temporal context windows seems to be dependent on idiosyncratic characteristics of each dataset.

\subsection{Model Interpretation}

\subsubsection{Technique Attention Visualization}
Figure~\ref{fig:beta_class} shows the attention attribution of each technique by a given technique. Tonal classes relied more on lower-to-middle layers, suggesting stronger dependence on local timbral cues. In contrast, articulation and legato classes placed more weight on middle layers, indicating the need for broader temporal context. These results show that playing technique benefited from different levels of MERT representation, supporting the use of learnable layer weighting in each expert.

\subsubsection{Technique-Dependent Expert Routing}
We next examined the extent each expert attended to a given technique in terms of their gating distributions on the VioTech120 test set. As shown in Figure~\ref{fig:gate_class}, tonal techniques tended to favor experts with shorter temporal context windows (<160ms), whereas articulation and legato favored experts with longer windows. We note that around 70\% of the weight was assigned to experts with temporal context windows above 320 ms when attending to legato, and a shift in the proportion of expert assignment from 320ms to 640ms for sustained vs bow-change frames. Further, we observed that pizzicato was differentiated among other tonal techniques by increased weighting at the 320 ms window.

\subsubsection{UMAP Analysis}
\begin{figure}[t]
    \centering
    \includegraphics[width=0.9\linewidth, trim={0cm 0cm 0cm 0.8cm}, clip]{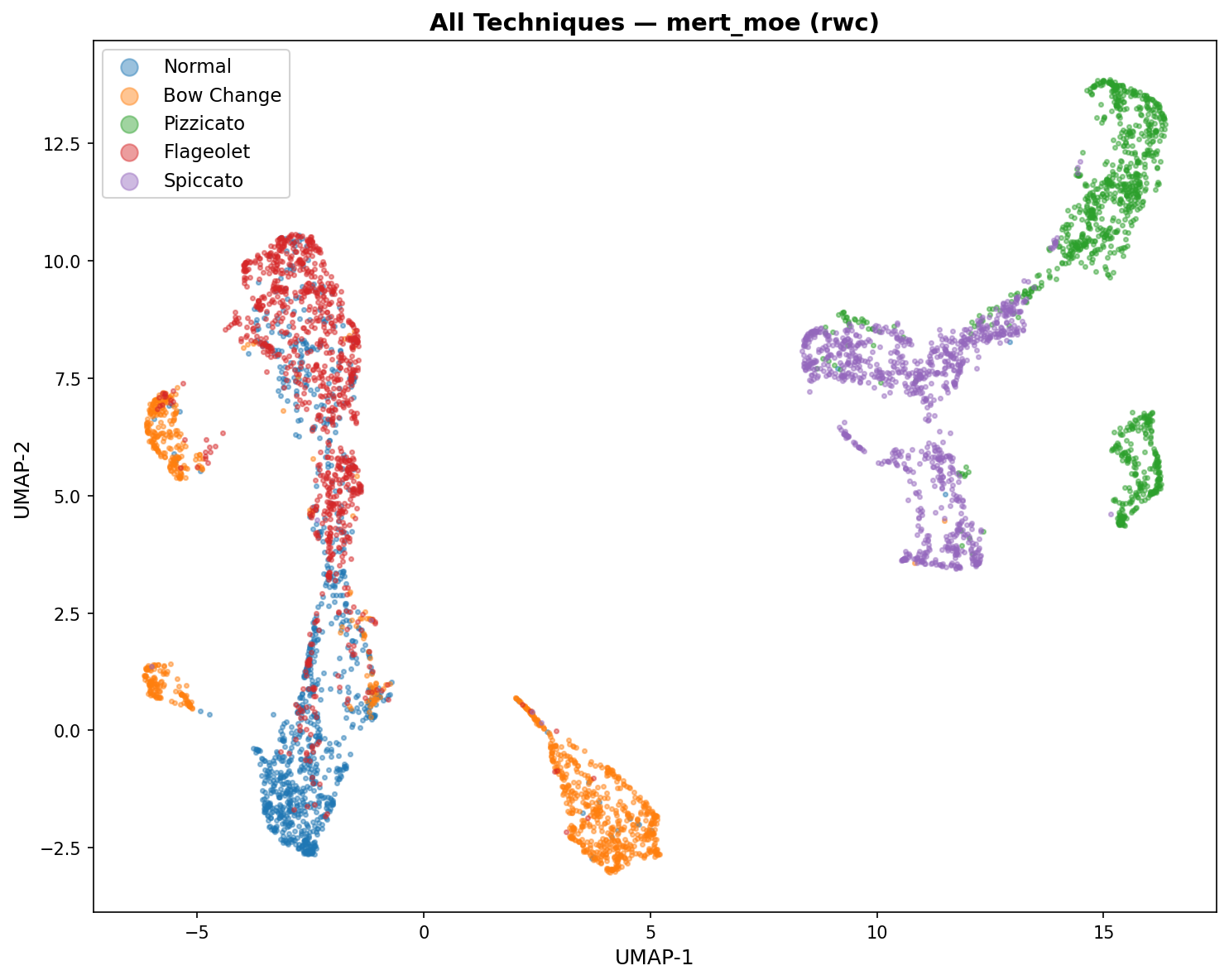}
    \caption{UMAP visualization of frame-level embeddings of each playing technique from the RWC dataset \cite{goto2003rwc}.}
    \label{fig:umap}
\end{figure}

We visualize 1,000 frame-level embeddings for each playing technique on the RWC dataset using UMAP (Figure~\ref{fig:umap}). The figure shows clear class-dependent structure. Pizzicato and spiccato form separated regions, reflecting their distinctive transient characteristics. Normal and flageolet partially overlap, suggesting shared bowed-string acoustic properties. Bow-change frames form separate clusters from normal frames, indicating that the model captures temporal cues related to changes in bowing. These results suggest that the learned embeddings encode meaningful differences among violin playing techniques.

\section{Conclusion}

Here, we presented VioTech, a frame-level violin playing technique detection model that addresses the limitations of note-level pipelines relying on onset and offset segmentation. Our mixture-of-experts approach further captured technique patterns across different temporal ranges to enable state-of-the-art transcription performance. We also introduced a manually labeled high-quality violin technique dataset from real performance recordings, together with an annotation interface for future violin technique and fingering labeling. That our model learned technique-dependent representations across feature levels and temporal scales further suggests that explicit temporal-scale specialization is a promising direction for violin technique detection.
\clearpage

\section{Acknowledgments}
This research work is partly supported by the Academia Sinica Grand Challenge Program (GCP) Seed Grant (AS–GCS–112–M07) and National Science and Technology Council (NSTC 114-2221-E-001-011-MY2). The author thanks the anonymous reviewers for their helpful comments that improved the quality of the paper.

\section{AI Usage Statement}
This work made use of a LLM to assist with 
data preprocessing, code implementation, organizing the 
related work discussed, and writing assistance for the paper. All LLM-assisted content was 
reviewed and verified by the authors, who take full responsibility 
for the accuracy and originality of the paper, including its 
citations.

\bibliography{ISMIR2026_VioTech}

\end{document}